\documentclass[12pt]{iopart}
\begin{document}

\title[Mother templates for gravitational wave chirps]{Mother templates for gravitational wave chirps}
\author{B.S. Sathyaprakash}
\address{Department of Physics and Astronomy, Cardiff University, Cardiff, UK}

\begin{abstract}
Templates used in a search for binary black holes and neutron stars  in 
gravitational wave interferometer
data will have to be computed on-line since the computational storage
and retrieval costs for the template bank are too expensive.
The conventional dimensionless variable $T=(c^3/Gm)t,$ where $m$ is the
total mass of a binary, in the time-domain and a not-so-conventional 
velocity-like variable $v=(\pi Gm f)^{1/3}$ in the Fourier-domain,
render the phasing of the waves independent of the total mass of
the system enabling the construction of {\it mother templates} that
depend only on the mass ratio of a black hole binary.  Use of such 
mother templates in a template bank will bring about a reduction 
in computational costs up to a factor of 10 and a saving on 
storage by a factor of 100.
\end{abstract}


Gravitational waves from binary black holes during the last few seconds
of their inspiral and merger are good candidates for a first direct observation
by interferometric gravitational wave detectors presently under construction.
The computational costs of generating and storing templates used in 
a search for these waves is quite expensive. Conventionally, one has used
a two-parameter family of templates corresponding to the two masses $m_1$
and $m_2$ of the component stars in a binary (equivalently, 
one has also used  the total mass $m=m_1+m_2$ and the (symmetric) mass ratio 
$\eta=m_1m_2/m^2$ or the Newtonian and post-Newtonian {\it chirp times,} 
etc.). However, as is well-known, general relativity allows us to use
a dimensionless time-variable $T=(c^3/Gm)t$ in terms of which all dynamical
equations become independent of the total mass but depend only on the
mass ratio $\eta.$ Therefore, relativists need
only study a one-parameter family of black hole binaries of different 
mass ratios; the conclusion they draw will be applicable for binaries
of different total masses. Of course, in doing so one has to use an
appropriate re-scaling of all physical scales.  

In gravitational wave data analysis one has so-far not taken 
advantage of this scaling since: (1) one deals with the detector data acquired
in the real time $t$ and not the adimensional time $T$ and (2)
one requires for the purpose data analysis samples 
$h_k\equiv h(t_k),$ $t_k\equiv k\Delta t,$ that are equally spaced in an apriori
chosen time-interval $\Delta t,$ corresponding to a sampling rate 
$f_s=1/\Delta t.$ It is obvious that the samples 
$h_K\equiv h(T_K),$ $T_K\equiv K\Delta T,$ 
equally spaced in $T$ yield the required samples $h_k$ only 
for a binary whose total mass is an integral multiple
of $m=c^3 M_c/G,$ where $M_c$ is some characteristic mass.
In this article we show that a one-parameter family of samples 
$h_K(\eta),$ supplemented 
with a simple linear interpolation, are sufficiently accurate for the 
construction of a two-dimensional search template bank $h_k(m,\eta).$
This brings about a saving on computational costs by a factor of up to
10 and storage costs by a factor 100. 

In the rest of this article we work with units $c=G=1.$
\section{Time-domain waveform}
Post-Newtonian (PN) theory computes the gravitational wave (GW) flux
${F}(v)$ emitted, and the (dimensionless) relativistic binding energy $E(v)$
of, a compact binary as expansions in the gauge independent 
velocity $v$ in the system. For a binary consisting of two non-spinning
black holes, of masses $m_1$ and $m_2,$ in quasi-circular orbit 
about each other, the flux and energy functions are both known
to order $v^5$ (i.e. 2.5 PN) beyond the quadrupole approximation
\cite{Blanchet etal 95,Blanchet 96}:
\begin{eqnarray}
\fl F(v) & = & \frac{32\eta^2 v^{10}}{5} 
      \left [ 1 - \left(\frac{1247}{336} + \frac{35}{12}\eta\right) v^2 
       + 4\pi v^3 \right .  \nonumber \\
\fl  & + & \left(- \frac{44711}{9072} 
       +   \left . \frac{9271}{504}\eta + 
            \frac{65}{18}\eta^2 \right ) v^4 
      -   \left(\frac{8191}{672} + \frac{535}{24}\eta\right) \pi v^5 
 + O(v^6) \right ], 
\label{eq:flux}
\end{eqnarray} 
\begin{eqnarray}
\fl E(v) = -\frac{\eta v^2}{2} \left [ 1 - \left ( \frac{9+\eta}{12} \right ) v^2 
     - \left ( \frac{81-57\eta+\eta^2}{24} \right ) v^4 
 + O(v^6) \right ]. 
\label{eq:energy}
\end{eqnarray}
Gravitational waves are dominantly emitted at twice the orbital frequency
of the system. The GW frequency is related to the invariant velocity $v$ via 
$v= (\pi m f_{\rm GW})^{1/3}.$ In the quasi-circular,
adiabatic approximation one uses the energy balance between the flux of
gravitational waves lost from the system and the rate of decay of
the binding energy of the system: $F(v(t))=-m(dE/dt)$.

The energy balance equation can be used to set up a pair of
coupled, non-linear, ordinary differential equations (ODEs) \cite{DIS3} to
compute the orbital phase evolution $\varphi(t)$ of the binary 
during the adiabatic regime:
\begin{equation}
\frac{d\varphi}{dt} - \frac{v^3}{m}=0, \ \
\frac{dv}{dt}  + \frac{F(v)}{mE'(v)}=0,
\label{eq:ODE}
\end{equation}
where we have made use of the relation between the orbital frequency and
the GW frequency, viz, $f_{\rm orb}(t) = \dot \varphi(t)/(2\pi) = 
f_{\rm GW}(t)/2 = v^3(t)/(2\pi m)$ and $E'(v)$ denotes the $v$-derivative 
of $E(v):$ $E'(v)=dE/dv.$  The above differential equations allow us to compute 
a two-parameter family of phasing formulas $\varphi_k^{m,\eta},$ 
$t_k= k \Delta t,$ corresponding to the parameters $(m,\eta),$ of the
binary that helps us to construct search templates.  Until now this 
has been the method of computing and storing templates. 
However, the cost of generating and storing templates in this way 
is quite expensive.

In the {\em restricted} PN approximation,
where the amplitude of the waveform is kept to the lowest PN 
order, while the phase is expanded to the highest PN
order known,  the GW radiation
emitted by a binary at a distance $r$ from Earth and sensed
by an interferometric antenna, is described by the waveform 
\begin{equation}
h(t) = \frac {4C\eta m}{r}  v^2(t) \cos[\phi(t)],   
\label{eq:time-domain signal}
\end{equation}
where $\phi(t)=2\varphi(t)$ is the gravitational wave phase.
$C$ is a constant that takes values in the range $[0,1]$ 
depending on the relative orientation of the source and the antenna.
It has an r.m.s (averaged over all orientations and wave polarisations) 
value of $2/5.$

Let us introduce a dimensionless time-variable $T=t/m.$ 
In terms of this new variable
the differential equations in equation (\ref{eq:ODE}) take the form
\begin{equation}
\frac{d\varphi}{dT} - v^3=0, \ \
\frac{dv}{dT}  + \frac{F(v)}{E'(v)}=0.
\label{eq:ODE2}
\end{equation}
The solutions of these two differential equations would  yield a phasing
formula $\varphi(T; \eta)$ that depends, unlike the phasing formula
$\varphi(t; m, \eta),$  on only one parameter, namely
the mass ratio $\eta.$ The advantage of this new phasing formula is that
for the purpose of data analysis it is sufficient to
create and store a finely sampled one-parameter family 
of {\it mother templates}
$\varphi_K^\eta,$ where $T_K=K\Delta T.$ The phasing 
formula for a binary of a given total mass $m$ and mass ratio 
$\eta,$ namely $\varphi_k^{m,\eta},$ can be computed from
the mother template, supplemented by a linear interpolation, without 
having to solve the ODEs all over again:
\begin{equation}
\varphi_k^{m,\eta} = (1-x) \varphi_K^\eta + x\varphi_{K+1}^\eta,\ \ 
K= \left [k/m \right ],\ \ x = k/m-K. 
\end{equation}
where for any real number $f,$ $[f]$ denotes the integer closest to, 
but smaller than, $f.$ This way of computing templates saves on computational 
costs by a factor of between 5 (for $T-$approximants) and 10 
(for $P$-approximants), depending on the signal model, and storage costs 
by a factor of 100 (the number of templates in the `$m$-coordinate').

In solving the differential equations above 
one must specify the initial conditions, namely the orbital phase and velocity 
at a chosen instant of time, say the instant of coalescence. These will, 
of course, depend on the total mass of the system. However, by choosing 
the instant of coalescence to be zero one can make any explicit 
dependence on the total mass vanish. 

\section{Frequency-domain waveform}
We shall show in this Section that the result we found in the 
previous Section for the time-domain phasing formula holds good in the
frequency domain too. 

The stationary phase approximation to the Fourier transform of 
the waveform in equation (\ref{eq:time-domain signal}) is given by
\cite{thorne 87,dhurandhar et al,sathyaprakash and dhurandhar 91}
\begin{eqnarray}
\tilde h(f) & \equiv & \int_{-\infty}^\infty h(t) e^{2\pi i f t} dt
\nonumber \\
& = & \frac{4C\eta m}{r} \int_{-\infty}^\infty \frac{v^2(t)} {2} 
\left [e^{i\psi^+_f(t)} + e^{i\psi^-_f(t)} \right] dt
\nonumber \\
& \simeq & \frac {2C\eta m}{r}  \frac{v^2(t_f)}{\sqrt{\dot{f}_{\rm GW}(t_f)}}   
e^{i [\psi_f(t_f) - \pi/4]},
\label{eq:frequency-domain signal}
\end{eqnarray}
where $\psi^\pm_f = 2\pi f t \pm \phi(t),$ 
an overdot denotes a derivative w.r.t. $t$ and
$t_f$ is the stationary point of the Fourier phase, $\psi^-_f(t)$ when
$f\ge 0$ and  $\psi^+_f(t)$ when $f \le 0,$
namely $\dot\psi_f^\pm(t_f)=0.$ Indeed, $t_f$ turns out to be the instant 
when the GW frequency $f_{\rm GW}$ is numerically equal
to the Fourier frequency $f:$ $f=\dot \phi(t_f)/(2\pi) = f_{\rm GW}(t_f).$ 
One can solve for $t_f$ by inverting the PN expansion of $f_{\rm GW}(t).$ 

On substituting for the stationary point $t_f$, 
and consistently using the available PN expansions 
of the flux and energy functions one gets the usual stationary
phase approximation (uSPA) to the inspiral signal 
\begin{equation}
\fl \psi_f(t_f) = 2\pi f t_C - \phi_C + 2 \int_{v_f}^\infty dv
(v_f^3-v^3)\frac{E'(v)}{F(v)}=
2\pi f t_C - \phi_C + \Psi(f;m,\eta)
\label{eq:phasing}
\end{equation}
where $v_f \equiv v(t_f) = (\pi m f)^{1/3}$.
$t_C$ is the time of coalescence and 
$\phi_C$ is the phase at $t=t_C,$ both of which will be set to zero. 
Just as in the time-domain, the frequency-domain phasing is also
given, in the SPA, by a pair of coupled, non-linear, ODEs:
\begin{equation}
\frac{d\psi(f)}{df} - 2\pi t(v_f) = 0, \ \
\frac{dt(v_f)}{df} + \frac{m (\pi m)^{1/3}}{3} \frac{E'(v_f)}{F(v_f)} 
\frac{1}{f^{2/3}} = 0.
\label {eq:frequency-domain ode}
\end{equation}
From the computational point of view, solving the ODE's above is
more efficient than computing the phase algebraically using
equation (\ref{eq:phasing}).
For most binaries the uSPA is sufficiently accurate. One, therefore,
generates the signal directly in the Fourier domain using 
equation (\ref{eq:frequency-domain ode}).  This is a lot quicker than solving 
the ODEs in equation ~(\ref{eq:ODE}) and then Fourier transforming. 
As in the time-domain, one can reduce the seemingly two-parameter
family of differential equations in equation ~(\ref{eq:frequency-domain ode})
to a one-parameter family by using $v=(\pi m f)^{1/3}$ as a Fourier 
variable instead of $f$ and a dimensionless `time function' 
$T(v) = t(v)/m.$ In that case the resulting equations are 
\begin{equation}
\frac{d\psi(v)}{dv} - 6v^2 T(v) = 0, \ \
\frac{dT(v)}{dv} + \frac{E'(v)}{F(v)} = 0.
\label {eq:frequency-domain ode2}
\end{equation}


\section{Explicit fast frequency-domain signal models}

We shall now give explicit formulas for the frequency-domain signal
in $v$-representation for $T$-approximants. As for $P$-approximants
it is best to use the differential equations in 
equation ~(\ref{eq:frequency-domain ode2}).

$\Psi(f)$ can be found by substituting in equation ~(\ref{eq:phasing})
for the flux and energy from
equations (\ref{eq:flux}) and (\ref{eq:energy}), re-expanding the
ratio $E'(v)/F(v)$ and integrating term-by-term. The resulting
Fourier phase can be  conveniently expressed as 
$\Psi(f;m,\eta) = \sum_0^{4} \Psi_k(f) \tau_k,$ where $\Psi_k$ are
functions only of the Fourier frequency and
$\tau_k$ are the so-called (dimensionless)
PN {\it chirp times} given in terms of the binary mass parameters 
\begin{eqnarray}
\fl \tau_0 & = & \frac{5}{256\eta v_0^5}, \  \tau_1=0,\ 
\tau_2 = \frac{5}{192\eta v_0^3}   
\left ( \frac{743}{336} + \frac{11}{4} \eta\right ), 
\tau_3 = \frac{\pi}{8\eta v_0^2},   \nonumber \\ 
\fl \tau_4 & = & \frac{5}{128\eta v_0}   
\left ( \frac{3058673}{1016064} + \frac{5429}{1008} \eta +   
\frac{617}{144}\eta^2 \right),
\label {eq:chirp times}
\end{eqnarray}   
with $v_0=(\pi m f_0)^{1/3}$,
$f_0$ is a fiducial frequency  (e.g., the lower cutoff of the
antenna response)
and the $\Psi_k$ are given, 
in terms of the scaled frequency  $\nu \equiv f/f_0$, by:
\begin{equation}   
\Psi_0 = \frac{6}{5\nu ^{5/3}},\   
\Psi_2 = \frac{2}{\nu },\   
\Psi_3 = \frac{-3}{\nu ^{2/3}},\   
\Psi_4 = \frac{6}{\nu ^{1/3}}.
\label {eq:algebraic phases}
\end{equation}   
The amplitude  in (\ref{eq:frequency-domain signal}) 
is a simple power-law of the form $f^{-7/6}$ \cite{thorne 87}.

The new representation that can speed up signal generation is
obtained by working in the Fourier domain with the post-Newtonian 
expansion parameter $v.$ Introduce a velocity-like variable 
$v.$ Indeed, $v$ is nothing but the parameter $v_f$ in equation (\ref{eq:phasing}). 
Use $v$ as the Fourier variable instead of $f.$ 
On substituting $f=v^3/(\pi m),$  and
setting the instant of coalescence $t_C$ and the phase at $\phi_C$ to zero,
we find 
\begin{equation}
\Psi(f) = \sum_{k=0}^4 \theta_k v^{k-5},
\end{equation}
where the {\it chirp parameters} $\theta_k$ are given by
\begin{eqnarray}
\fl \theta_0 & = & \frac{3}{128\eta}, \  
\theta_1=0,\  
\theta_2=\frac{5}{96\eta}
\left ( \frac{743}{336} + \frac{11}{4} \eta\right ), 
\theta_3 = -\frac{3\pi}{8\eta}, \nonumber \\
\fl \theta_4 & =& \frac{15}{64\eta}   
\left ( \frac{3058673}{1016064} + \frac{5429}{1008} \eta +   
\frac{617}{144}\eta^2 \right).
\end{eqnarray}   
The Fourier transform in equation \ref{eq:frequency-domain signal} 
can now be written as
\begin{equation}
\tilde h(f) = \frac {Cm^2}{r}\sqrt{\frac{5\pi\eta }{384 v^7}} 
\exp \left [i \sum_{k=0}^4 \theta_k v^{k-5} - i\pi/4 \right],
\label{eq:velocity-domain}
\end{equation}
Note that in this form the phase depends only on the mass ratio $\eta$ and
not on the total mass $m$\cite{footnote1}.
One can take advantage
of this feature in lowering the computational costs in
generating templates in the following manner: One computes a set of look-up
Tables, or mother templates, $\Psi(v; \eta)$ vs $v$ for different 
values $\eta.$ The Fourier
phase $\Psi(f; m,\eta),$ for a binary of mass $m$ and mass ratio $\eta,$
can simply be read off from the appropriate look-up table:
$\Psi(f; m, \eta) = \Psi(v=(\pi m f)^{1/3}; \eta).$ Moreover, the cost of
storing a one-parameter set of tables of 
$\Psi(v; \eta)$ should be several orders of magnitude lower
than the cost of storing a two-parameter family of Tables
$\Psi(f; m, \eta).$

\section{Practical Considerations}

The codes for generating mother templates have been implemented and 
tested for accuracy and are available for a free-download \cite{codes}. 
We find that the waveforms generated by the interpolation method
agree well with the waveforms directly computed using the ODEs. 
However, certain practical
matters should be kept in mind while using these codes. Firstly, it is
most accurate to work with either of the time- or frequency-domain
phases, $\varphi(T)$ or $\Psi(v),$ respectively, rather
than the corresponding waveforms. In other words, one must compute
and store a mother phasing formula and {\it never} a mother waveform.
Interpolating waveforms is much harder, and more inaccurate, than
the phase of a waveform, which is a monotonic smooth function. Secondly, the 
initial velocity $v_0$ of a mother template required in solving the 
differential equations for the phasing $\varphi(t)$ must correspond 
to the binary of lowest total mass $m_{\rm min}$ in a search: 
$v_0=(\pi m_{\rm min}f_0)^{1/3},$
where $f_0$ is the lower frequency cutoff of a detector.
Thirdly, if the sampling interval in real time is $\Delta t$ then the
sampling interval $\Delta T$ in adimensional time 
should be chosen equal to $\Delta t/m_{\rm max},$ where $m_{\rm max}$ is
the maximum mass of in a search, in order for the interpolating formula to 
work best. The last two points can be summarised as: {\it the smallest 
velocity in a mother template should correspond to the lightest binary
in a search and the adimensional sampling rate should correspond to 
the heaviest binary.}
This means that if the real sampling rate is $f_t,$ then in a search for 
binaries of masses $m<10 M_\odot$ one must choose a sampling rate 
$f_T \simeq 5 \times 10^{-2} (f_t/1kHz) (m_{\rm max}/10 M_\odot)$
\cite{caviate1}.

Finally, the integration of the ODEs must be terminated at, or 
just before, the time when the system reaches the last stable orbit 
defined by the velocity at which the energy function has an
extremum, that is $E'(v_{\rm lso})=0.$ The right had side of the
equation for velocity evolution in equation (\ref{eq:ODE2}) diverges at
this point and integration cannot be continued beyond $v=v_{\rm lso}.$
At certain post-Newtonian orders and approximations the velocity evolution
close to, but well before, the last stable orbit is not monotonic; these
correspond to the case where the flux function $F(v)$ has a zero before
the last stable orbit velocity. In those cases one must stop the 
integration before reaching $dv/dT=0,$ to avoid the code from crashing.

\section{Conclusions}

The use of an adimensional time, well-known in the literature on black
hole binaries, in the time-domain, or a velocity-like parameter in the
Fourier-domain, allows us to introduce a one-parameter family of
{\it mother} templates which significantly reduces the computational 
costs of generating
and storing binary inspiral search templates that will be used in the
analysis of data from the up-coming ground-based gravitational wave 
interferometers.  Mother templates of binary inspiral waveforms
would also speed-up data analysis simulations that require a large number
of waveforms of different masses.

It is not clear at the present time what implications 
this representation will have for the actual choice of search templates. 
It is conceivable that one will
be able to define a `$v$-transform' of the data so that the problem
becomes a one-parameter search for binary black holes with different
mass ratios. However, let us not forget that such a scheme will only
be able to detect those signals whose instant of coalescence is $t_C=0;$
in other words only those that coalesce at the end of a given data segment.
One will, therefore, have to explicitly search for binaries with every possible 
instant of coalescence, as opposed to the presently planned scheme wherein one
takes advantage of the fact that one can search for (almost) all instants
of coalescence, in one-go, by working in the Fourier-domain. 
A new scheme, therefore, is unlikely to significantly change the computational
costs of a search, although the method may have certain advantages over the
conventional approach. Whether or not an alternative method based on the
mother templates introduced in this work is yet to be seen. 

Mother templates that are independent of the total mass of a binary will be
available at all post-Newtonian orders, in the effective one-body
approach \cite{buonanno and damour 00,DIS3} and in the 
case of eccentric and spinning binaries too. Their use should make 
it easier for us to incorporate new parameters, such as eccentricity 
and spins, in our search templates to detect compact binaries in 
gravitational wave data.

\ack
I am thankful to LIGO for a visiting associateship at Caltech where most 
of this work was done. For helpful discussions I am indebted to 
R. Balasubramanian, K. Blackburn, A. Lazzarini and T. Prince.


\section*{References}
\begin{thebibliography}{10}
\bibitem{Blanchet etal 95}
   L. Blanchet, T. Damour, B.R. Iyer, C.M. Will and A. Wiseman,  
   Phys. Rev. Lett.  {\bf 74}, 3515 (1995)
\bibitem{Blanchet 96}
   L. Blanchet, Phys. Rev. {\bf D54}, 1417 (1996).
\bibitem {DIS3}
   T. Damour, B.R. Iyer and B.S. Sathyaprakash, {\it A comparison of search
   templates for gravitational waves from binary inspiral} gr-qc/0010009 (2000).
\bibitem {thorne 87} 
   K.S. Thorne, {\it Gravitational Radiation,} in S.W. Hawking
   and W. Isreal (eds), {\it 300 Years of Gravitation,} (Cambridge University
   Press, Cambridge).
\bibitem {codes} The codes are available from URL: 
   www.astro. cf.ac.uk/pub/B.Sathyaprakash. 
\bibitem {dhurandhar et al}
   S.V. Dhurandhar, J. Watkins and B.F. Schutz, {\it Fourier transform
of a coalescing binary signal} (unpublished preprint, 1989); see 
\protect{\cite{sathyaprakash and dhurandhar 91}} for a derivation.
\bibitem {sathyaprakash and dhurandhar 91}
   B.S. Sathyaprakash and S.V. Dhurandhar, Phys. Rev. {\bf D44}, 3819 (1991).
\bibitem {buonanno and damour 00}
   A. Buonanno and T. Damour, Phys. Rev. {\bf D62,} 064015 (2000).
\bibitem {footnote1}
   {Although the amplitude does depend on the total mass, it is 
   irrelevant in template generation.} 
\bibitem {caviate1}
The sampling rate in adimensional time is a factor 
$m_{\rm max}/m_{\rm min}$ larger than the minimal required sampling rate.
This could be a large factor requiring very large memory to hold each 
mother template. However, one can circumvent the problem
by working with a small number of mother templates for each mass ratio.
\end{thebibliography}
\end{document}